\documentclass[aps,prd,reprint,groupedaddress,amsmath]{revtex4-2}
\usepackage[colorlinks,linkcolor=blue,hyperfootnotes=true]{hyperref}
\usepackage{graphicx}% Include figure files
\usepackage{dcolumn}% Align table columns on decimal point
\usepackage{bm}% bold math
\usepackage{footnote}
\usepackage{extarrows}
\usepackage{longtable}
\usepackage{threeparttable}
\usepackage{tabularx}

\begin{document}

%\preprint{}

%\title{The Gravitational Bending in Acoustic Schwarzschild Black Hole and Gauss-Bonnet Theorem}
\title{Curvatures, Photon Spheres and Black Hole Shadows}


\author{Chen-Kai Qiao}
\email{Email: chenkaiqiao@cqut.edu.cn}
%\thanks{chenkaiqiao@cqut.edu.cn}
\affiliation{College of Science, Chongqing University of Technology, Banan, Chongqing, 400054, China}


\date{\today}

\begin{abstract}
In a recent work \href{https://doi.org/10.1103/PhysRevD.106.L021501}{PRD {\bf 106}, L021501 (2022)}, a new geometric approach is proposed to obtain the photon sphere (circular photon orbit) and the black hole shadow radius. In this approach, photon spheres and the black hole shadow radius are determined using geodesic curvature and Gaussian curvature in the optical geometry of black hole spacetimes. However, the calculations in \href{https://doi.org/10.1103/PhysRevD.106.L021501}{PRD {\bf 106}, L021501 (2022)} only restricted to a subclass of static and spherically symmetric black holes with spacetime metric $g_{tt} \cdot g_{rr}=-1$, $g_{\theta\theta}=r^{2}$ and $g_{\phi\phi}=r^{2}\sin^{2}\theta$. In this work, we extend this approach to more general spherically symmetric black holes (with spacetime metric $ds^{2}=g_{tt}dt^{2}+g_{rr}dr^{2}+g_{\theta\theta}d\theta^{2}+g_{\phi\phi}d\phi^{2}$). Furthermore, it can be proved that our results from the geometric approach are completely equivalent to those from conventional approach based on effective potentials of test particles. 

\ \ 

Geodesic Curvature, Spherically Symmetric Black Hole
\end{abstract}


\maketitle

\section{Introduction \label{section1}}

Black holes are massive compact objects in our universe, which play central roles in Einstein's general relativity and other gravity theories. In the past decades, black holes have attracted a large amount of interest in high-energy physics, astrophysics, astronomy and cosmology. Significantly important information on gravitation, galaxies, thermodynamics and quantum effects in curved spacetime can be revealed from the explorations of black holes \cite{Bekenstein1973,Hawking1974,Hawking1975,Witten1998,Virbhadra2000,Ferrarese2000,Ryu2006,Black Hole Physics,Carr2021}. Recently, huge progresses in black hole physics have been witnessed. The gravitational wave signals from binary black hole mergers were detected by LIGO and Virgo Collaborations \cite{LIGO2016,LIGO2016b}. The high-resolution images of supermassive black holes in M87 galaxy and Sgr A* were successfully captured by Event Horizon Telescope (EHT) \cite{ETH2019a,ETH2022}.

The photon sphere (also called the circular photon orbit, or light ring) and shadow radius are important quantities in black hole physics. The particle motions, gravitational lensing, optical imaging and other aspects of black hole can be studied from these quantities. Since the observation of high-precision black hole images in the M87 galaxy and Sgr A* by the EHT Collaboration, photon spheres and black hole shadows have become extremely attractive topics in physics and astronomy. Conventionally, the photon sphere and black hole shadow radius can be calculated from the effective potential of test particles moving in black hole spacetime \cite{Hioki2008,Pugliese2011,Johannsen2013,Guo2020,Carroll,Hartle,Perlick2022,Raffaelli2021,Gan2021,WangMZ2022}. However, other approaches to photon spheres, which are brought up mostly in virtue of topological and geometric techniques, also emerged in the past years \cite{Gibbons1993,Virbhadra2001,Cunha2017,Cunha2018,Cunha2020,Ghosh2021,LiuYX2019}. Recently, P. V. P. Cunha, E. Berti and C. A. R. Herdeiro demonstrated the existence of photon spheres using the topological index of vector fields in black hole spacetime, and they constrained the number of photon spheres by a similar method \cite{Cunha2017,Cunha2018,Cunha2020}. Later, S. -W. Wei, Y. -X. Liu and R. B. Mann studied the shape and boundary length of black hole shadow for Kerr spacetime using the local curvature radius of shadow's boundary curves and the Gauss-Bonnet theorem in the celestial coordinate plane \cite{LiuYX2019}. These new approaches, which served as supplementary of the conventional approach, could give us new insights on black hole physics and spacetime geometry. A number of important questions about photon spheres and black hole shadows can be investigated from these new approaches.

\begin{table}
	\caption{Distinguishing features of the geometric approach developed by Qiao and Li for a subclass of spherically symmetric black holes in a recent work (reference \cite{Qiao}).}
	\label{table1}
	\vspace{2mm}
	\begin{ruledtabular}
		\begin{tabular}{lcc}
			\multicolumn{2}{c}{Geometric approach to photon spheres and black hole shadows}
			\\
			\hline
			Geometry & Optical geometry of black hole spacetime 
			\\
			\hline
			Key quantities & Gaussian curvature $\mathcal{K}(r)$ 
			\\
			& Geodesic curvature $\kappa_{g}(r)$       
			\\
			\hline
			photon sphere  & $\kappa_{g}(r)=0$ 
			\\
			unstable photon sphere & $\kappa_{g}(r)=0$ and $\mathcal{K}(r)<0$ 
			\\
			stable photon sphere & $\kappa_{g}(r)=0$ and $\mathcal{K}(r)>0$ 
			\\
		\end{tabular}
	\end{ruledtabular}
\end{table}

Recently, C. K. Qiao and M. Li proposed a new geometric approach to obtain the photon spheres and black hole shadows for a subclass of static and spherically symmetric black holes \cite{Qiao}.  In this approach, the photon spheres and black hole shadows are reflected by the geometric properties for optical geometry of black hole spacetime. Especially, the Gaussian curvature and geodesic curvature in the equatorial plane of optical geometry can completely determine the stable and unstable photon spheres. The distinguishing features of this approach are summarized in table \ref{table1}. Besides, it is also proved that the geometric approach developed by Qiao and Li is completely equivalent to the conventional approach based on effective potential of test particles moving in black hole spacetime \cite{Qiao}. Furthermore, P. V. P. Cunha, C. A. R. Herdeiro and J. P. A. Novo modified this geometric approach to study the timelike circular geodesics in the Jocabi geometry of black hole spacetimes \cite{Cunha2022}.

However, the original calculations by Qiao and Li in reference \cite{Qiao} only restricted in a subclass of static and spherically symmetrical black hole with spacetime metric $g_{tt} \times g_{rr} = -1$, $g_{\theta\theta}=r^{2}$ and $g_{\phi\phi}=r^{2}\sin^{2}\theta$. It is necessary to see that whether the same approach can be applied to more general black holes. In this work, we give an extension of this geometric approach to more general spherically symmetric black holes with the spacetime metric from $ds^{2}=g_{tt}dt^{2}+g_{rr}dr^{2}+g_{\theta\theta}d\theta^{2}+g_{\phi\phi}d\phi^{2}$. Our results shows that the photon sphere $r=r_{ph}$ exactly vanishes its geodesic curvature in optical geometry, namely $\kappa_{g}(r=r_{ph})=0$. The negative (or positive) Gaussian curvature in equatorial plane of optical geometry would indicate the correspond photon spheres are unstable (or stable). These conclusions are exactly the same with the conclusions in reference \cite{Qiao}, which have been summarized in table \ref{table1}.  Furthermore, in the present work, we also prove that the calculations using the geometric approach for more general spherically symmetric black holes are completely equivalent to the conventional approach based on effective potential of test particles.

This work is organized in a following way. Section \ref{section1} gives the backgrounds and motivations of the present work. Section \ref{section2} gives a brief introduction of optical geometry for black holes. In section \ref{section3}, the Gauss curvature and geodesic curvature are described. The photon spheres and black halo shadow radius for general spherically symmetrical black holes are calculated and analyzed in section \ref{section4} and section \ref{section5}. Section \ref{section6} provides explicit results on photon spheres and black hole shadow radius for some typical examples of black hole solutions. The equivalence between the geometric approach (based on Gauss curvature and geodesic curvature in optical geometry) and the conventional approach (based on effective potential of test particles) is demonstrated in section \ref{section7}. The summary and perspectives are given in section \ref{section8}. In this work, the natural unit $G=c=1$ is used.

\section{Optical Geometry of Black Hole Spacetime \label{section2}}

This section gives an introduction of the optical geometry of black hole spacetime. The optical geometry is a powerful tool to study the motions of photons (or other massless particles which travel along null geodesics) in the gravitational field  \cite{Abramowicz1988,Gibbons2008,Gibbons2009,Werner2012}. The optical geometry can be obtained from spacetime geometry $ds^{2}=g_{\mu\nu}dx^{\mu}dx^{\nu}$, which is a four dimensional Lorentz manifold, by imposing the null constraint $d\tau^{2}=-ds^{2}=0$. 
\begin{equation}
	\underbrace{ds^{2} = g_{\mu\nu}dx^{\mu}dx^{\nu}}_{\text{Spacetime Geometry}}
	\ \ \overset{d\tau^{2}=-ds^{2}=0}{\Longrightarrow} \ \ 
	\underbrace{dt^{2} = g^{\text{OP}}_{ij}dx^{i}dx^{j}}_{\text{Optical Geometry}}
	\label{optical geometry}
\end{equation}
Further, if we consider particle motions in the equatorial plane, a two dimensional manifold can be constructed from the optical geometry.
\begin{equation}
	\underbrace{dt^{2} = g^{\text{OP}}_{ij}dx^{i}dx^{j}}_{\text{Optical Geometry}}
	\ \ \overset{\theta=\pi/2}{\Longrightarrow} \ \ 
	\underbrace{dt^{2}=\tilde{g}^{\text{OP-2d}}_{ij}dx^{i}dx^{j}}_{\text{Optical Geometry (Two Dimensional)}}
	\label{optical geometry2}
\end{equation}
In this two dimensional manifold, a number of elegant theorems in surface theory and differential geometry could provide powerful techniques to study the particle motions near black holes. In the geometric approach developed by Qiao and Li, the analysis and calculations on photon spheres / circular photon orbits and black hole shadows are implemented in the two dimensional optical geometry.

The optical geometry attracted a number of interests in recent years, and many aspects of black hole physics can be analyzed and studied using optical geometry \cite{Abramowicz1988,Gibbons2008,Gibbons2009}. For instance, G. W. Gibbons and M. C. Werner developed an approach to calculate the gravitational deflection angle using the Gauss-Bonnet theorem in optical geometry \cite{Gibbons2008}. For stationary black hole spacetime, the photon orbits (which are along lightlike / null geodesics in a four dimensional Lorentz manifold), become spatial geodesics and minimize the stationary time $t$ when they are transformed into optical geometry. The stationary time coordinate $t$ plays the role of arc-length parameter / spatial distance parameter in optical geometry, and $\delta[\int_{\gamma}dt]=0$ is satisfied along the photon orbits. This conclusion can be viewed as the generalization of the renowned Fermat's principle in the curved manifold  \cite{Gibbons2009,Werner2012,footnote1}. %In the geometric approach, both the photon sphere and black hole shadow radius are analyzed and calculated in the equatorial plane of optical geometry. 

The properties of optical geometry strongly depend on symmetries of the gravitational field and black hole spacetime. For a spherically symmetric black hole, its optical geometry gives a Riemannian manifold \cite{Abramowicz1988,Gibbons2008,Gibbons2009}. Consider a spherically symmetric black hole spacetime with the metric
\begin{equation}
	ds^{2} = g_{tt}dt^{2}+g_{rr}dr^{2}+g_{\theta\theta}d\theta^{2}+g_{\phi\phi}d\phi^{2}
\end{equation} 
its optical geometry is obtained from the null constraint $d\tau^{2}=-ds^{2}=0$, which eventually gives a three dimensional Riemannian manifold
\begin{equation}
	dt^{2} = g_{ij}^{\text{OP}}dx^{i}dx^{j}
	= -\frac{g_{rr}}{g_{tt}}\cdot dr^{2} - \frac{g_{\theta\theta}}{g_{tt}} \cdot d\theta^{2} - \frac{g_{\phi\phi}}{g_{tt}} d\phi^{2} 
	\label{optical geometry3}
\end{equation}
When studying the photon spheres and black hole shadows, one can always restrict this optical geometry to the equatorial plane $\theta=\pi/2$ without loss of generality. 
\begin{equation}
	dt^{2}=\tilde{g}^{\text{OP-2d}}_{ij}dx^{i}dx^{j}
          = - \frac{g_{rr}}{g_{tt}} \cdot dr^{2} - \frac{\overline{g}_{\phi\phi}}{g_{tt}} \cdot d\phi^{2} 
\end{equation}
where $\tilde{g}^{\text{OP-2d}}_{ij}$ is the optical metric restricted in the two dimensional equatorial plane, and the notation $\overline{g}_{\phi\phi}$ is defined as $\overline{g}_{\phi\phi}=g_{\phi\phi}(\theta=\pi/2)$. Besides, for static and spherically symmetric black holes, another equivalent way to define an optical geometry is from the conformal transformation of spacetime and extracting its spatial part \cite{Gibbons2009}.

For rotational / axi-symmetric black holes, the correspond optical geometry is a Randers-Finsler manifold \cite{Werner2012,Ono2017,Jusufi2018,Jusufi2018b}. In these cases, we should consider the standard rotational black hole metric
\begin{equation}
	ds^{2} = g_{tt}dt^{2}+2g_{t\phi}dtd\phi+g_{rr}dr^{2}+g_{\theta\theta}d\theta^{2}+g_{\phi\phi}d\phi^{2}
\end{equation} 
the optical geometry can be obtained in a similar way by imposing the null constraint $d\tau^{2}=-ds^{2}=0$. The arc length / spatial distance in optical geometry eventually reduces to
\begin{eqnarray}
	dt & = & \sqrt{-\frac{g_{rr}}{g_{tt}} \cdot dr^{2} - \frac{g_{\theta\theta}}{g_{tt}} \cdot d\theta^{2} + \frac{g_{t\phi}^{2}-g_{tt}g_{\phi\phi}}{g_{tt}^{2}} \cdot d\phi^{2}} \nonumber
	\\
	   &   & - \frac{g_{t\phi}}{g_{tt}}\cdot d\phi
\end{eqnarray}
This gives a class of Renders-Finsler manifolds. The Renders-Finsler geometry is the extension of Riemannian geometry \cite{ChernSS}, and the arc-length / spatial distance can be separated into two parts
\begin{equation}
	dt = \sqrt{\alpha_{ij}dx^{i}dx^{j}} + \beta_{i}dx^{i} \label{Renders geometry}
\end{equation}
Here, the first part $\alpha_{ij}$ is a Riemannian metric, and the second part $\beta=\beta_{i}dx^{i}$ is a one-form which measures the departure of this Renders-Finsler geometry from the Riemannian geometry $dt^{2}=\alpha_{ij}dx^{i}dx^{j}$. The above Renders-Finsler geometry recovers the Riemannian geometry if and only if $\beta=0$.

In this work, we only deal with spherically symmetric black holes, whose optical geometry is a Riemannian manifold. Therefore, in the present work, both Gaussian curvature and geodesic curvature are calculated using formulas in Riemannian geometry. The investigations on photon spheres and black hole shadows for rotational black holes, whose optical geometry gives a Renders-Finsler manifold, are left for future studies.

\section{Gaussian Curvature and Geodesic Curvature \label{section3}}

There are several basic quantities which describe the geometric properties of optical geometry. In the geometric approach proposed by Qiao and Li, the Gaussian curvature and geodesic curvature in the equatorial plane of optical geometry play central roles in determining the photon spheres of black holes. 

In the surface theory and differential geometry, the Gaussian curvature is the intrinsic curvature of a two dimensional surface $S$, which measures how far this surface is from being flat intrinsically. The geodesic curvature is the curvature of a curve $\gamma(s)$ lived in this surface, which measures how far this curve is from being a geodesic in surface $S$ \cite{Carmo1976,Berger,Berger2}.  If $\gamma$ is a geodesic curve in surface $S$, its geodesic curvature $\kappa_{g}(\gamma)$ automatically vanishes. Both geodesic curvature and Gaussian curvature are intrinsic geometric quantities of the two dimensional surface $S$, and they can be calculated from the intrinsic metric of surface $S$ naturally. 

If we assign the curving linear coordinates $(x_{1},x_{2})$ for a two-dimensional surface $S$ such that its intrinsic metric is expressed as
\begin{equation}
	ds^{2}=g_{11}dx_{1}^{2}+g_{22}dx_{2}^{2} \label{intrinsic metric}
\end{equation}
for an arbitrary curve $\gamma=\gamma(s)=(x_{1}(s),x_{2}(s))$ living in the surface $S$ with arc-length parameter / spatial distance parameter $s$, its geodesic curvature can be calculated through the following equation \cite{Carmo1976,ChernWH}
\begin{eqnarray}
	\kappa_{g}(\gamma) & = & \frac{d\alpha}{ds} 
	- \frac{1}{2\sqrt{g_{22}}}\frac{\partial \ log(g_{11})}{\partial x_{2}} \cos\alpha \nonumber
	\\
	&   & + \frac{1}{2\sqrt{g_{11}}}\frac{\partial \ log(g_{22})}{\partial x_{1}} \sin\alpha
\end{eqnarray}
Here, $\alpha$ is the angle between tangent vector of $\gamma(s)$ and coordinate axis $x_{1}$. Since the geodesic curvature measures how far a curve is from being a geodesic curve in two dimensional surface $S$, the vanishing of the geodesic curvature indicates that $\gamma=\gamma(s)$ is a geodesic curve in this surface.
\begin{subequations}
\begin{eqnarray}
	\kappa_{g}(\gamma) = 0 
	& \Leftrightarrow & 
	\nabla_{T} T \bigg|_{\gamma=\gamma(s)} = 0 
	\\
	\kappa_{g}(\gamma) = 0 
	& \Leftrightarrow & 
	\bigg[ \frac{d^{2}x_{i}}{ds^{2}}-\Gamma_{jk}^{i}\frac{dx_{j}}{ds}\frac{dx_{k}}{ds} \bigg]_{\gamma=\gamma(s)} = 0
\end{eqnarray}
\end{subequations}
Here, $T$ is the tangent vector along this curve $\gamma$ in the two-dimensional surface. Particularly, if $S$ is a flat two-dimensional space, one can assign the osculating circle at each point $p$ along the curve $\gamma$, such that the osculating circle tangent to this curve at a given point $p$. Under this circumstance, the inverse of the geodesic curvature gives the local curvature radius at this point, $R(\gamma)=1/\kappa_{g}(\gamma)$. Recently, some works have established the relationship between black hole shadows and local curvature radius in the celestial coordinate plane of black holes \cite{LiuYX2019}. However, in a general curved space, it is difficult (not always possible) to assign the ``osculating circle'' at each point along a curve, and the concept of a local curvature radius becomes subtle.

Using the intrinsic metric in equation (\ref{intrinsic metric}), the Gaussian curvature of two dimensional surface $S$ can be calculated through the expression \cite{Carmo1976,ChernWH}
\begin{eqnarray}
	\mathcal{K} & = & \mathcal{K}_{1}\mathcal{K}_{2}
	= \frac{R_{1212}}{g_{11}g_{22}-(g_{12}^{2})} \nonumber
	\\
	& = & 
	-\frac{1}{\sqrt{g_{11}g_{22}}}
	\bigg[
	\frac{\partial}{\partial x_{2}} \bigg( \frac{1}{\sqrt{g_{22}}} \frac{\partial\sqrt{g_{11}}}{\partial x_{2}}  \bigg)
	+ \frac{\partial}{\partial x_{1}} \bigg( \frac{1}{\sqrt{g_{11}}} \frac{\partial\sqrt{g_{22}}}{\partial x_{1}}  \bigg)
	\bigg] \nonumber
	\\ \label{Gauss-Curvature1}
\end{eqnarray}
where $\mathcal{K}_{1}$, $\mathcal{K}_{2}$ are principle curvatures of surface $S$, and $R_{1212}$ is the Riemannian curvature tensor of this surface. In this section, we have adopted the notations in surface theory, where the covariant and contravariant coordinates are not distinguished. 
 
\section{Photon Sphere \label{section4}}

In this section, we use the geometric approach in reference \cite{Qiao} to obtain the stable and unstable photon spheres for more general spherically symmetric black holes. The original calculations in reference \cite{Qiao} are restricted to a subclass of static and spherically symmetric black holes with the spacetime metric $g_{tt} \cdot g_{rr} = -1$, $g_{\theta\theta}=r^{2}$ and $g_{\phi\phi}=r^{2}\sin^{2}\theta$. The general spacetime metric for spherically symmetric black holes can be written as
\begin{equation}
	ds^{2}=g_{tt}(r)dt^{2}+g_{rr}(r)dr^{2}+g_{\theta\theta}(r)d\theta^{2}+g_{\phi\phi}(r,\theta)d\phi^{2}
	\label{spacetime metric}
\end{equation}
Because of the spherical symmetry, when discussing photon spheres and other photon orbits, we can always restrict the particle motions in the equatorial plane without loss of generality. The optical geometry restricted in the equatorial plane $\theta=\pi/2$ is given by 
\begin{eqnarray}
	dt^{2} & = & \tilde{g}^{\text{OP-2d}}_{ij}dx^{i}dx^{j} \nonumber
	\\
	& = & \tilde{g}^{\text{OP-2d}}_{rr}dr^{2} + \tilde{g}^{\text{OP-2d}}_{\phi\phi}d\phi^{2} \nonumber
	\\
	& = & - \frac{g_{rr}(r)}{g_{tt}(r)} \cdot dr^{2} - \frac{\overline{g}_{\phi\phi}(r)}{g_{tt}(r)} \cdot d\phi^{2}
\end{eqnarray}
where $\overline{g}_{\phi\phi}(r)=g_{\phi\phi}(r,\theta=\pi/2)$. The photon sphere could be determined by the Gaussian curvature and geodesic curvature in the equatorial plane of optical geometry. The optical geometry of black hole spacetime has the following interesting property. The photon orbits, which are along null geodesic curve in spacetime geometry (which is four dimensional Lorentz geometry), become spatial geodesic curves when they are transformed into optical geometry. In other words, the null geodesic curve $\gamma=\gamma(\tau)$ in spacetime geometry $ds^{2}=g_{\mu\nu}dx^{\mu}dx^{\nu}$ maintains geodesic in the optical geometry $dt^{2}=g^{\text{OP}}_{ij}dx^{i}dx^{j}$. %Actually, this observation can be viewed as the generalization of Fermat's principle in curved stationary spacetime \cite{Gibbons2009,Werner2012,footnote1}.

The photon sphere $r=r_{ph}$ near the black hole, which is a geodesic curve in spacetime geometry, becomes spatial geodesic curve in the equatorial plane of optical geometry. Thus the geodesic curvature of photon sphere $r=r_{ph}$ vanishes naturally \cite{Carmo1976,ChernWH,Li2021}
\begin{eqnarray}
	& & \ \ 
	\kappa_{g}(r=r_{ph}) = \frac{1}{2\sqrt{\tilde{g}^{\text{OP-2d}}_{rr}}} 
	\frac{\partial \big[ log(\tilde{g}^{\text{OP-2d}}_{\phi\phi})\big]}{\partial r} \bigg|_{r=r_{ph}} = 0 \nonumber 
	\\
	& \Rightarrow & \ 
	\bigg\{
	  \frac{1}{2\sqrt{-\frac{g_{rr}(r)}{g_{tt}(r)}}} \cdot
	  \frac{\partial}{\partial r}
	  \bigg[ \log\bigg(-\frac{\overline{g}_{\phi\phi}(r)}{g_{tt}(r)}\bigg) \bigg]
	\bigg\}_{r=r_{ph}}
	= 0 \nonumber 
	\\
	& \Rightarrow & \ 
	\bigg[
	  \frac{g_{tt}(r)}{2\overline{g}_{\phi\phi}(r)} \cdot \sqrt{-\frac{g_{tt}(r)}{g_{rr}(r)}} 
	  \cdot \frac{\partial}{\partial r} \bigg( \frac{\overline{g}_{\phi\phi}(r)}{g_{tt}(r)} \bigg)
	\bigg]_{r=r_{ph}}
	= 0 \nonumber
	\\ \label{geodesic cuurvature general}
\end{eqnarray}
The solution of this equation gives the exact position of photon spheres. In this way, the photon sphere / circular photon orbit in spherically symmetric black holes can be obtained analytically. Furthermore, for the study of photon spheres, we can always carry out calculations in regular regions of black hole spacetimes (in which singularities are excluded). Therefore we assume that the spacetime metric is non-degenerate in the calculations, namely $g_{tt}(r)$, $g_{rr}(r)$, $g_{\theta\theta}(r)$ and $g_{\phi\phi}(r,\theta)$ are nonzero. In such cases, the geodesic curvature condition for photon spheres can be further simplified as
\begin{eqnarray}
	\kappa_{g}(r=r_{ph}) = 0 & \Rightarrow & \ 
	\frac{\partial}{\partial r} \bigg( \frac{\overline{g}_{\phi\phi}(r)}{g_{tt}(r)} \bigg)
	\bigg|_{r=r_{ph}}
    = 0 \label{geodesic cuurvature general simplified}
\end{eqnarray}
Actually, the optical geometry is usually defined outside the black hole horizons, where $g_{tt}(r)>0$ and the optical metric in equation (\ref{optical geometry3}) always makes sense. In such cases, the coordinate $t$ can be regarded as time measured by an observer.

Our results can provide as the natural generation of the geometric approach in reference \cite{Qiao} for more general spherically symmetric black holes. To recover the results in reference \cite{Qiao}, we restrict the spacetime metric to a subclass of the spherically symmetric metric such that $g_{tt}(r)=-f(r)$, $g_{rr}(r)=1/f(r)$, $g_{\theta\theta}(r)=r^{2}$, $g_{\phi\phi}(r,\theta)=r^{2}\sin^{2}\theta$, then the above photon sphere condition in equation (\ref{geodesic cuurvature general}) reduces to
\begin{eqnarray}
	\kappa_{g}(r=r_{ph}) 
	& = & 
	\bigg[
	  \frac{g_{tt}(r)}{2\overline{g}_{\phi\phi}(r)} \cdot \sqrt{-\frac{g_{tt}(r)}{g_{rr}(r)}} 
	  \cdot \frac{\partial}{\partial r} \bigg( \frac{\overline{g}_{\phi\phi}(r)}{g_{tt}(r)} \bigg)
	\bigg]_{r=r_{ph}} \nonumber
	\\
	& = &
	\bigg[
	  \frac{1}{2\overline{g}_{\phi\phi}(r)} \cdot \sqrt{-\frac{g_{tt}(r)}{g_{rr}(r)}} \cdot \frac{\partial \overline{g}_{\phi\phi}(r)}{\partial r} \nonumber
	\\
	&   & \ \ 
	  + \frac{1}{2\sqrt{-g_{tt}(r) \cdot g_{rr}(r)}} \cdot \frac{\partial g_{tt}(r)}{\partial r}
	\bigg]_{r=r_{ph}} \nonumber
	\\
	& = &
	\bigg[ \frac{f(r)}{r}-\frac{1}{2}\cdot\frac{\partial f(r)}{\partial r} \bigg]_{r=r_{ph}} 
	= 0 \label{geodesic cuurvature simplified}
\end{eqnarray}
This is exactly the same as the equation (5) in reference \cite{Qiao}. 

According to the geometric approach proposed by Qiao and Li, the stability of photon spheres near black holes are completely determined by the Gaussian curvature in the equatorial plane optical geometry. In reference \cite{Qiao}, this conclusion is obtained by analyzing bound photon orbits near the photon spheres, as well as other geometric and topological constraint (especially the Hadamard theorem in differential geometry and topology). The argument in the Hadamard theorem only constrains the Gaussian curvature, irrespective of the specific metric form. Therefore, the same analysis can be exactly applicable to a more general spherically-spherical black hole spacetime in equation (\ref{spacetime metric}). Finally, the following criterion is obtained to determine the stability of photon spheres / circular photon orbits.
\begin{eqnarray}
	\mathcal{K}(r) < 0 & \Rightarrow & \text{The photon shere $r=r_{ph}$ is unstable}  \nonumber \\
	\mathcal{K}(r) > 0 & \Rightarrow & \text{The photon sphere $r=r_{ph}$ is stable} \nonumber
\end{eqnarray}
The connections between Gaussian curvature, stability of photon spheres and the Hadamard theorem in differential geometry are described in Appendix \ref{appendix1}. For spherically symmetric black holes, the Gaussian curvature of optical geometry restricted in the equatorial plane can be calculated through \cite{Carmo1976,ChernWH}
\begin{widetext}
\begin{eqnarray}
	\mathcal{K} & = & -\frac{1}{\sqrt{\tilde{g}^{\text{OP-2d}}}} 
	\bigg[
	\frac{\partial}{\partial \phi} \bigg( \frac{1}{\sqrt{\tilde{g}^{\text{OP-2d}}_{\phi\phi}}} \frac{\partial\sqrt{\tilde{g}^{\text{OP-2d}}_{rr}}}{\partial \phi}  \bigg)
	+ \frac{\partial}{\partial r} \bigg( \frac{1}{\sqrt{\tilde{g}^{\text{OP-2d}}_{rr}}} \frac{\partial\sqrt{\tilde{g}^{\text{OP-2d}}_{\phi\phi}}}{\partial r}  \bigg)
	\bigg]  \nonumber
	\\
	& = & 
	- \frac{1}{\sqrt{\frac{g_{rr}(r)}{g_{tt}(r)}\cdot \frac{\overline{g}_{\phi\phi}(r)}{g_{tt}(r)}}}
	  \bigg\{
	    \frac{\partial}{\partial\phi}
	    \bigg[ 
	      \frac{1}{\sqrt{-\frac{\overline{g}_{\phi\phi}(r)}{g_{tt}(r)}}} \cdot
	      \frac{\partial}{\partial \phi}
	      \bigg( \sqrt{-\frac{g_{rr}(r)}{g_{tt}(r)}} \bigg)
	    \bigg] 
	    +\frac{\partial}{\partial r}
	     \bigg[
	       \frac{1}{\sqrt{-\frac{g_{rr}(r)}{g_{tt}(r)}}} \cdot
	       \frac{\partial}{\partial r}
	       \bigg( \sqrt{-\frac{\overline{g}_{\phi\phi}(r)}{g_{tt}(r)}} \bigg)
	     \bigg]
	  \bigg\} \nonumber
	\\
	& = & \frac{g_{tt}(r)}{\sqrt{g_{rr}(r)\cdot \overline{g}_{\phi\phi}(r)}}
	      \cdot \frac{\partial}{\partial r}
	      \bigg[
	        \frac{g_{tt}(r)}{2\sqrt{g_{rr}(r)\cdot \overline{g}_{\phi\phi}(r)}}
	        \cdot \frac{\partial}{\partial r} \bigg( \frac{\overline{g}_{\phi\phi}(r)}{g_{tt}(r)} \bigg)
	      \bigg] \label{Gauss cuurvature general}
\end{eqnarray}
Here, $\tilde{g}^{\text{OP-2d}}=det(\tilde{g}^{\text{OP-2d}}_{ij})$ is the determinant of the two dimensional optical metric.

To recover the results in reference \cite{Qiao}, one can also restrict the spacetime metric as $g_{tt}(r)=-f(r)$, $g_{rr}(r)=1/f(r)$, $g_{\theta\theta}(r)=r^{2}$ and $g_{\phi\phi}(r,\theta)=r^{2}\sin^{2}\theta$. In these conditions, the Gaussian curvature of two dimensional optical geometry reduces to
\begin{eqnarray}
	\mathcal{K}(r) 
	& = & 
	\frac{g_{tt}(r)}{\sqrt{g_{rr}(r)\cdot \overline{g}_{\phi\phi}(r)}}
	\cdot \frac{\partial}{\partial r}
	\bigg[
	  \frac{g_{tt}(r)}{2\sqrt{g_{rr}(r)\cdot \overline{g}_{\phi\phi}(r)}}
	  \cdot \frac{\partial}{\partial r} \bigg( \frac{\overline{g}_{\phi\phi}(r)}{g_{tt}(r)} \bigg)
	\bigg] \nonumber
	\\
	& = & 
	\frac{1}{2} f(r) \cdot \frac{d^{2}f(r)}{dr^{2}}
	- \bigg[ \frac{1}{2} \cdot \frac{df(r)}{dr} \bigg]^{2} \label{Gauss curvature simplified}
\end{eqnarray}
which is exactly the same as equation (6) in reference \cite{Qiao}. In this way, we have demonstrated that the present work on photon spheres of more general spherically symmetric black holes can be viewed as a natural extension of the geometric approach proposed in reference \cite{Qiao}.
\end{widetext}

\section{Black Hole Shadow Radius \label{section5}} 

\begin{figure}
	\includegraphics[width=0.525\textwidth]{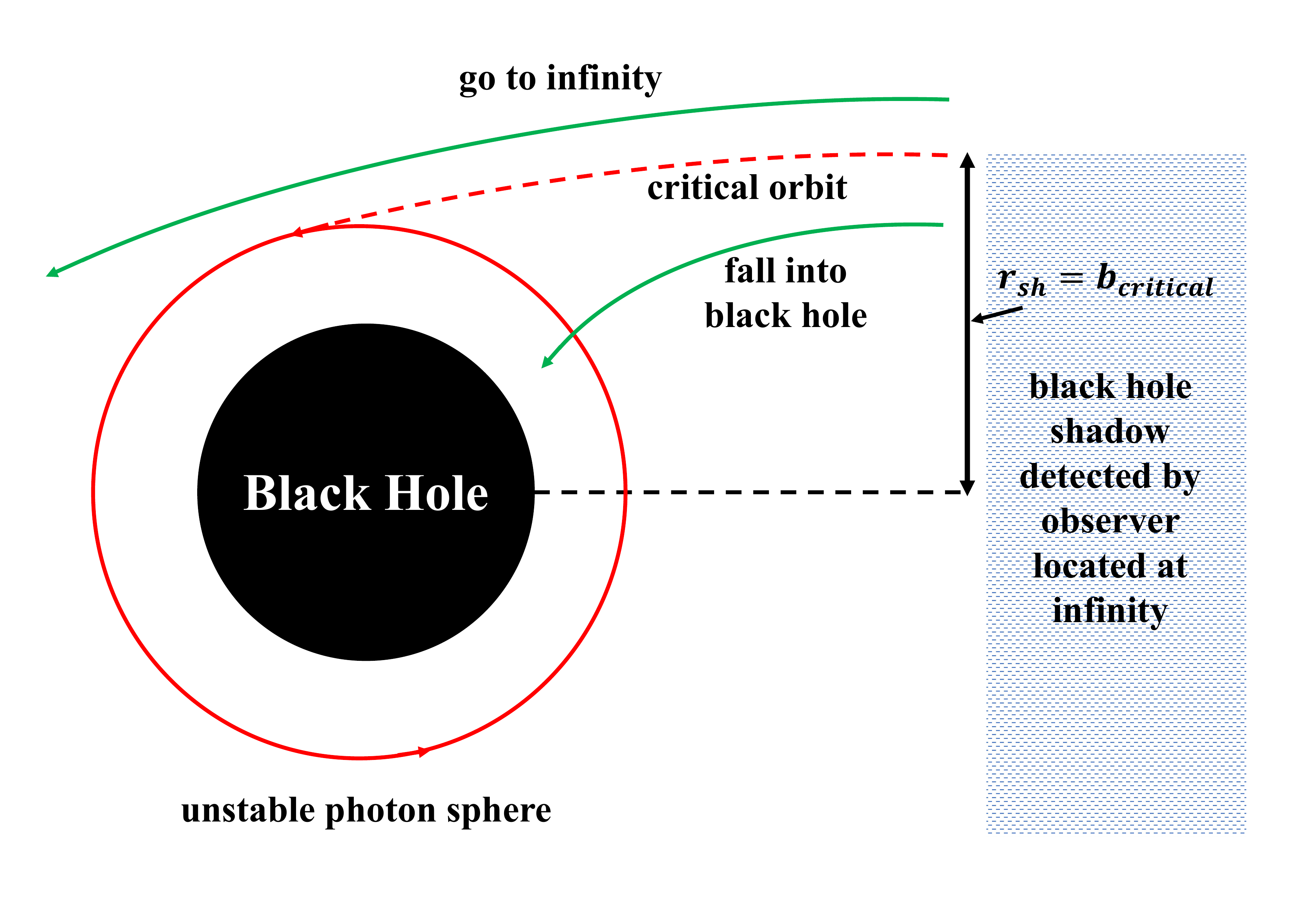}
	\caption{This figure shows the connection between black hole shadows and unstable photon spheres. The region covered by black denotes the interior of a spherically symmetric black hole. The light beam emitted from infinity with impact parameter $b=b_{\text{critical}}$ would reach the unstable photon sphere exactly. The light beam emitted with impact parameter $b<b_{\text{critical}}$ would fall into the black hole.}
	\label{figure3}
\end{figure}

The black hole shadow is the dark silhouette of black hole image in a bright background. The size and shape of black hole shadow depends not only on black hole parameters, but also on the position of observers \cite{Perlick2022}. In this section, we concentrate ourselves on the idealized situation where observer is located at infinity, and there are no light sources between the observer and the central black hole. In this case, the radius of black hole shadow detected by observer is just the critical value of impact parameter $b_{\text{critical}}$. In the gravitational field, light beams emitted from infinity with impact parameter $b=b_{\text{critical}}$ would reach the unstable photon sphere exactly, as illustrated in figure \ref{figure3}

In this section, we follow the procedure in reference \cite{Qiao} to derive the black hole shadow radius $r=r_{sh}$ for more general spherically symmetric black holes. According to the conventional definition, the impact parameter can be expressed as \cite{Hartle,Carroll}
\begin{equation}
	b \equiv \bigg|\frac{L}{E}\bigg|
	%=   \bigg|
	%\frac{r^{2}\sin^{2}\theta \cdot d\phi/d\lambda}{f(r)\cdot dt/d\lambda}
	%\bigg|
\end{equation}
where $E$ and $L$ are the conserved energy and conserved angular momentum, respectively 
\begin{subequations}
	\begin{eqnarray}
		E & \equiv & - K_{t} \cdot T = - g_{tt}(r)\cdot\frac{dt}{d\lambda}  \label{conserved energy}
		\\
		L & \equiv & K_{t} \cdot T = g_{\phi\phi}(r,\theta) \cdot \frac{d\phi}{d\lambda} \label{conserved angular momnetum}
	\end{eqnarray}
\end{subequations}
The $K_{t}=\partial/\partial t$ and $K_{\phi}=\partial/\partial\phi$ are two Killing vector fields in static and spherically symmetric black hole spacetime, and $T=d/d\lambda$ is the tangent vector of photon orbits ($\lambda$ is arbitrary affine parameter along null geodesics). Since the impact parameter $b \equiv L/E$ 
is a conserved quantity along geodesics, we can calculate the critical impact parameter $b_{\text{critical}}$ at any specific point along the photon orbits. A simple and convenient choice are the points in the unstable photon sphere / unstable circular photon orbit $r=r_{\text{unstable}}$ 
\begin{eqnarray}
	b_{\text{critical}} & = & 
	\bigg| \frac{L}{E} \bigg|_{r=r_{\text{unstable}}} \nonumber
	\\
	& = & \bigg| -\frac{g_{\phi\phi}(r,\theta) \cdot d\phi/d\lambda}{g_{tt}(r) \cdot dt/d\lambda} \bigg|_{r=r_{\text{unstable}}} \nonumber
	\\
	& = & \bigg| -\frac{\overline{g}_{\phi\phi}(r)}{g_{tt}(r)} \cdot \frac{d\phi}{dt} \bigg|_{r=r_{\text{unstable}}}
\end{eqnarray}
In the last line, $\theta=\pi/2$ has been used for orbits in the equatorial plane.

The radius of the black hole shadow detected by observer at infinity, which is the critical value of impact parameter ($r_{sh}=b_{\text{critical}}$), can be calculated in the optical geometry of black hole spacetime. In spacetime geometry, photon orbits are always along lightlike / null geodesics, and their tangent vector $T=d/d\lambda$ satisfies $T \cdot T = g_{\mu\nu} \frac{dx^{\mu}}{d\lambda} \frac{dx^{\nu}}{d\lambda} = 0$, with $\lambda$ to be any affine parameter of null geodesics. However, when photon orbits are transformed into the optical geometry, the photon orbits become spatial geodesics with vanishing geodesic curvature $\kappa_{g}(\gamma)=0$, as we have emphasized in previous sections. The stationary time coordinate $t$ in spacetime geometry exactly reduces to the arc-length parameter / spatial distance parameter in optical geometry. 
Furthermore, because the optical geometry of spherically symmetric black holes is a Riemannian geometry, the tangent vector of arbitrary photon orbits $T^{\text{OP}}=d/dt$ with respect to the arc-length parameter $t$ becomes a unit vector ($|T^{\text{OP}}|=1$) after normalization. Here, we restrict the optical geometry in the equatorial plane, and the tangent vector of photon orbits reduce to $T^{\text{OP-2d}}=(\frac{dr}{dt},\frac{d\phi}{dt})$. Furthermore, for unstable photon spheres / unstable circular photon orbits, we have the following relations
%\begin{widetext}
\begin{eqnarray}
	& & r=r_{\text{unstable}}=\text{constant} 
	\ \ \Rightarrow \ \ 
	\frac{dr}{dt} \bigg|_{r=r_{\text{unstable}}} = 0 \nonumber 
	\\
    & \Rightarrow & 
    |T^{\text{OP-2d}} \cdot T^{\text{OP-2d}}| = \bigg| \tilde{g}^{\text{OP-2d}}_{\phi\phi} \cdot  \frac{d\phi}{dt} \cdot \frac{d\phi}{dt} \bigg| \nonumber
    \\
    &             & \ \ \ \ \ \ \ \ \ \ \ \ \ \ \ \ \ \ \ \ \ \ \ 
    = \bigg| -\frac{\overline{g}_{\phi\phi}(r)}{g_{tt}(r)} \cdot \frac{d\phi}{dt} \cdot \frac{d\phi}{dt} \bigg|_{r=r_{\text{unstable}}}
    = 1 \nonumber
    \\
    & \Rightarrow & r_{sh} = b_{\text{critical}} 
    = \bigg| -\frac{\overline{g}_{\phi\phi(r)}}{g_{tt}(r)} \cdot \frac{d\phi}{dt} \bigg|_{r=r_{\text{unstable}}} \nonumber
    \\
    &             & \ \ \ \ \ \ \ \ \ \ \ \ \ \ \ \ \ \ 
    = \sqrt{-\frac{\overline{g}_{\phi\phi}(r)}{g_{tt}(r)}} \bigg|_{r=r_{\text{unstable}}} \label{black hole shadow}
\end{eqnarray}
%\end{widetext}
This is the analytical expression of the black hole shadow radius detected by observer located at infinity in an idealized situation. Recall $\overline{g}_{\phi\phi}(r)$ is exactly the metric component $g_{\phi\phi}(r,\theta)$ restricted in the equatorial plane $\theta=\pi/2$, it can be seen that the shadow radius calculated within our geometric approach is consistent with previous results based on conventional approach \cite{Perlick2022,Virbhadra2022}. 

\section{Some Examples \label{section6}}

%\begin{threeparttable*}
\setlength{\LTcapwidth}{480pt}
%\setlength{\textwidth}{500pt}
%\begin{widetext}
\begin{longtable*}{lc}
	\caption{The photon sphere / circular photon orbit $r=r_{ph}$ and black hole shadow radius detected by observer at infinity $r_{sh}=b_{\text{critical}}$. This table summarizes the results for Schwarzschild black hole, Reissner-Nordstr\"om black hole, Kottler black hole, Schwarzschild black hole in the quintessential field, the spherically symmetric black hole in conformal Weyl gravity, and the Schwarzschild-like black hole in bumblebee gravity. All quantities calculated using the geometric approach (based on geodesic curvature and Gaussian curvature in optical geometry) agree with the results obtained using the conventional approach (based on effective potential of test particles). }
	%\\
\label{table3}
\vspace{2mm}
%\begin{ruledtabular}
	%\begin{tabular}{lcc}
	%\begin{tabularx}{450pt}{lX}
	    \\
	    \hline
		Type of Black Hole & Schwarzschild Black Hole 
		\\
		\hline
		Spacetime Metric   & $ds^{2}=-(1-2M/r)\cdot dt^{2}+(1-2M/r)^{-1}\cdot dr^{2}+r^{2}\cdot d\theta^{2}+r^{2}\sin^{2}\theta\cdot d\phi^{2}$ 
		\\
		%\hline
		Geodesic Curvature & $\kappa_{g}(r) = 1/r-3M/r^{2}$ 
		\\
		%\hline
		Photon Sphere      & $\kappa_{g}(r=r_{ph})=0 \ \Rightarrow \ r_{ph}=3M$
		\\
		%\hline
		Gaussian Curvature & $\mathcal{K}(r)=-2M/r^{3}+3M^{2}/r^{4}<0$ 
		\\
		                   & (outside black hole horizon, $r>2M$)   \ \href{table footnote a}{$^{a}$}
		\\
		Stability of Photon Sphere & $r_{ph}=3M$ is an unstable photon sphere 
		\\
		Black Hole Shadow Radius   & $r_{sh}=b_{\text{critical}}=3\sqrt{3}M$ 
		\vspace{2mm}
		\\
		\hline
		Type of Black Hole & Reissner-Nordstr\"om Black Hole
		\\
		\hline
		Spacetime Metric & $ds^{2}=-(1-2M/r+Q^{2}/r^{2})\cdot dt^{2}+(1-2M/r+Q^{2}/r^{2})^{-1}\cdot dr^{2}+r^{2}\cdot d\theta^{2}+r^{2}\sin^{2}\theta\cdot d\phi^{2}$ 
		\\
		%\hline
		Geodesic Curvature & $\kappa_{g}(r) = 1/r-3M/r^{2}+2Q^{2}/r^{3} $
		\\
		%\hline
		Photon Sphere      & $\kappa_{g}(r=r_{ph})=0 \ \Rightarrow \  r_{ph}=(3M+\sqrt{9M^{2}-8Q^{2}})/2$
		\\
		%\hline
		Gaussian Curvature & $\mathcal{K}(r)=-2M/r^{3}+3(M^{2}+Q^{2})/r^{4}-6MQ^{2}/r^{5}+2Q^{4}/r^{6}<0$
		\\
		                   & (outside black hole horizon, $r>M+\sqrt{M^{2}-Q^{2}}$)
		\\
		Stability of Photon Sphere & $r_{ph}=(3M+\sqrt{9M^{2}-8Q^{2}})/2$ is an unstable photon sphere
		\\
		Black Hole Shadow Radius   &  $r_{sh}=b_{\text{critical}}=\frac{(3M+\sqrt{9M^{2}-8Q^{2}})^{2}}{\sqrt{8(3M^{2}-2Q^{2}+M\sqrt{9M^{2}-8Q^{2}})}}$
		\vspace{2mm}
		\\
		\hline
		Type of Black Hole & Kottler Black Hole 
		\\
		\hline
		Spacetime Metric & $ds^{2}=-(1-2M/r-\Lambda r^{2}/3)\cdot dt^{2}+(1-2M/r-\Lambda r^{2}/3)^{-1}\cdot dr^{2}+r^{2}\cdot d\theta^{2}+r^{2}\sin^{2}\theta\cdot d\phi^{2}$ 
		\\
		%\hline
		Geodesic Curvature & $\kappa_{g}(r) = 1/r-3M/r^{2}$ 
		\\
		%\hline
		Photon Sphere      & $\kappa_{g}(r=r_{ph})=0 \ \Rightarrow \ r_{ph}=3M$
		\\
		%\hline
		Gaussian Curvature & $\mathcal{K}(r)=-2M/r^{3}+3M^{2}/r^{4}+2\Lambda M/r-\Lambda/3<0$ \ (with $0<\Lambda<1/9M^{2}$)
		\\
		                   & (outside black hole horizon, namely $1-2M/r-\Lambda r^{2}/3>0$) \ \href{table footnote b}{$^{b}$}
		\\
		Stability of Photon Sphere & $r_{ph}=3M$ is an unstable photon sphere 
		\\
		Black Hole Shadow Radius   & $r_{sh}=b_{\text{critical}}=\frac{3\sqrt{3}M}{\sqrt{1-9\Lambda M^{2}}}$
		\vspace{2mm}
		\\
		\hline
		Type of Black Hole & Spherically Symmetric Black Hole in Conformal Weyl Gravity
		\\
		\hline
		Spacetime Metric   & $ds^{2}=-(1-3M\gamma-2M/r+\gamma r-k r^{2})\cdot dt^{2}+(1-3M\gamma-2M/r+\gamma r-k r^{2})^{-1}\cdot dr^{2}+r^{2}\cdot d\theta^{2}+r^{2}\sin^{2}\theta\cdot d\phi^{2}$  
		\\
		%\hline
		Geodesic Curvature & $\kappa_{g}(r) = (1-3M\gamma)/r-3M/r^{2}+\gamma/2$ 
		\\
		%\hline
		Photon Sphere      & $\kappa_{g}(r=r_{ph})=0 \ \Rightarrow \  r_{ph}= 3M-1/\gamma+\sqrt{1+9M^{2}\gamma^{2}}/\gamma$
		\\
		%\hline
		Gaussian Curvature & $\mathcal{K}(r)= -k(1-3M\gamma)-\gamma^{2}/4+6MK/r-3M\gamma/r^{2}-2M(1-3M\gamma)/r^{3}+3M^{2}/r^{4} <0$ \ (with $k>0$)  \href{table footnote c}{$^{c}$}
		\\
		                   & (outside black hole horizon, namely $1-3M\gamma-2M/r+\gamma r-k r^{2}>0$) 
		\\
		Stability of Photon Sphere & $r_{ph}= 3M-1/\gamma+\sqrt{1+9M^{2}\gamma^{2}}/\gamma$ is an unstable photon sphere ($k>0$)
		\\
		Black Hole Shadow Radius   & $r_{sh}=b_{\text{critical}}=\frac{3M-1/\gamma+\sqrt{1+9M^{2}\gamma^{2}}/\gamma}{\sqrt{\sqrt{1+9\gamma^{2}M^{2}}-2M/(3M-1/\gamma+\sqrt{1+9M^{2}\gamma^{2}}/\gamma)}-k(3M-1/\gamma+\sqrt{1+9M^{2}\gamma^{2}}/\gamma)^2}$
		\vspace{2mm}
		\\
		\hline
		Type of Black Hole &  Schwarzschild Black Hole in the Quintessential Field
		\\
		\hline
		Spacetime Metric   & $ds^{2}=-(1-2M/r-\alpha/r^{3\omega+1})\cdot dt^{2}+(1-2M/r-\alpha/r^{3\omega+1})^{-1}\cdot dr^{2}+r^{2}\cdot d\theta^{2}+r^{2}\sin^{2}\theta\cdot d\phi^{2}$ \ \href{table footnote d}{$^{d}$}
		\\
		%\hline
		Geodesic Curvature & $\kappa_{g}(r) = 1/r-3M/r^{2}-\alpha/2$ 
		\\
		%\hline
		Photon Sphere      & $\kappa_{g}(r=r_{ph})=0 \ \Rightarrow \  r_{ph}= (1-\sqrt{1-6M\alpha})/\alpha$
		\\
		%\hline
		Gaussian Curvature & $\mathcal{K}(r)= -\alpha^{2}/4+3M\alpha/r^{2}-2M/r^{3}+3M^{2}/r^{4} <0$ \ $(\omega=-2/3)$
		\\
		& (outside black hole horizon, $1/2\alpha-\sqrt{1-8M\alpha}/2\alpha<r<1/2\alpha+\sqrt{1-8M\alpha}/2\alpha$ ) \ \href{table footnote e}{$^{e}$}
		\\
		Stability of Photon Sphere & $r_{ph}= (1-\sqrt{1-6M\alpha})/\alpha$ is an unstable photon sphere \ $(\omega=-2/3)$
		\\
		Black Hole Shadow Radius   & $r_{sh}=b_{\text{critical}}=\frac{(1-\sqrt{1-6M\alpha})^{3/2}}{\alpha\sqrt{\sqrt{1-6M\alpha}+4M\alpha-1}}$ \ $(\omega=-2/3)$
		\vspace{2mm}
		\\
		\hline
		Type of Black Hole &  Schwarzschild-Like Black Hole in Bumblebee Gravity
		\\
		\hline
		Spacetime Metric   &  $ds^{2}=-(1-2M/r)\cdot dt^{2}+(l+1)\cdot(1-2M/r)^{-1}\cdot dr^{2}+r^{2}\cdot d\theta^{2}+r^{2}\sin^{2}\theta\cdot d\phi^{2}$ \ \href{table footnote f}{$^{f}$}
		\\
		Geodesic Curvature &  $\kappa_{g}(r) = (1/r-3M/r^{2})\cdot(1/\sqrt{1+l})$ 
		\\
		Photon Sphere      &  $\kappa_{g}(r=r_{ph})=0 \ \Rightarrow \ r_{ph}=3M$
		\\
		Gaussian Curvature &  $\mathcal{K}(r)=(1/1+l)\cdot(-2M/r^{3}+3M^{2}/r^{4})<0$ \ \  (with $1+l>0$)
		\\
		                   &  (outside black hole horizon, $r>2M$)
		\\
		Stability of Photon Sphere &  $r_{ph}=3M$ is an unstable photon sphere
		\\
		Black Hole Shadow Radius   &  $r_{sh}=b_{\text{critical}}=3\sqrt{3}M$
		\vspace{2mm}
		\\
		\hline
		\\
	%\end{tabularx}
%\end{ruledtabular}
%\begin{tablenotes}
	%\item[1] tablefootnote 1
	%\item[2] Note that this black hole has an event horizon $r=\frac{1-\sqrt{1-8M\alpha}}{2\alpha}$ and a cosmological horizon $r=\frac{1+\sqrt{1-8M\alpha}}{2\alpha}$ when $\omega=-\frac{2}{3}$.
%\end{tablenotes}
\multicolumn{2}{l}{\footnotesize{a. Note that optical geometry is usually defined outside the black hole horizon, where $g_{tt}(r)>0$ is satisfied and the optical metric in }}
\\
\multicolumn{2}{l}{\footnotesize{\ \ \ \ equation (\ref{optical geometry3}) always makes sense.}} 
\\
\multicolumn{2}{l}{\footnotesize{b. We assume the cosmological constant $0<\Lambda<\frac{1}{9M^{2}}$ in Kottler black hole.}}
\\
\multicolumn{2}{l}{\footnotesize{c. In conformal Weyl gravity, the parameter $M$ has the relation $M=\frac{\beta(2-3\beta\gamma)}{2}$ \cite{Mannheim1989}. Here, we choose the $k>0$ case in the calculation.}}
\\
\multicolumn{2}{l}{\footnotesize{d. In this spacetime metric, the $-1<\omega<-\frac{1}{3}$ is the quintessential field parameter, and $\alpha>0$ characterizes the scale of universe. Here, in }}
\\
\multicolumn{2}{l}{\footnotesize{\ \ \ \ order to get an analytical result and compare it with reference \cite{TaoJ2022,ZengXX2020}, we choose the parameter $\omega=-\frac{2}{3}$. In this case, the spacetime }}
\\
\multicolumn{2}{l}{\footnotesize{\ \ \ \ metric becomes $ds^{2}=-(1-2M/r-\alpha r)\cdot dt^{2}+(1-2M/r-\alpha r)^{-1}\cdot dr^{2}+r^{2}\cdot d\theta^{2}+r^{2}\sin^{2}\theta\cdot d\phi^{2}$.}}
\\
\multicolumn{2}{l}{\footnotesize{e. Note that this black hole has an event horizon $r=\frac{1-\sqrt{1-8M\alpha}}{2\alpha}$ and a cosmological horizon $r=\frac{1+\sqrt{1-8M\alpha}}{2\alpha}$ when $\omega=-\frac{2}{3}$.}}
\\
\multicolumn{2}{l}{\footnotesize{f. In the bumblebee gravity model, $l$ (with $l+1>0$) is a Lorentz-violating parameter \cite{Casana2018}.}}
\end{longtable*}
%\end{threeparttable*}
%\end{widetext}

According to the algorithm described in previous two sections, the photon spheres and shadow radius for an arbitrary spherically symmetric black hole with spacetime metric $ds^{2}=g_{tt}(r)dt^{2}+g_{rr}(r)dr^{2}+g_{\theta\theta}(r)d\theta^{2}+g_{\phi\phi}(r,\theta)d\phi^{2}$ can be calculated efficiently. In this section, we choose some spherically symmetric black holes as typical examples to carry out the calculations. They are the Schwarzschild black hole, Reissner-Nordstr\"om black hole, Kottler black hole, Schwarzschild black hole in the quintessential field, the spherically symmetric black hole in Weyl conformal gravity, and the Schwarzschild-like black hole in bumblebee gravity. These black hole solutions play significantly important roles in Einstein's general relativity and other gravity theories. The explicit results on the photon spheres and black hole shadow radius for these black holes are summarized in table \ref{table3}. From this table, it is clearly shown that all quantities calculated using the geometric approach (based on geodesic curvature and Gaussian curvature in the optical geometry of black holes) in the present work are consistent with the results obtained using conventional approach (based on effective potential of test particles) \cite{Perlick2022,Das2022,Perlick2018,Virbhadra2022,Li2021,TaoJ2022,ZengXX2020}. Particularly, the last example in table \ref{table3} illustrates the calculation with $g_{tt} \times g_{rr} \neq -1$. From this table, the photon sphere and black hole shadow radius in Schwarzschild-like black hole of bumblebee gravity are exactly the same as those in the conventional Schwarzschild black hole. This can be easily explained according to our analytical expression in equations (\ref{geodesic cuurvature general simplified}) and (\ref{black hole shadow}). Both the photon sphere $r=r_{ph}$ and the black hole shadow radius $r_{sh}=b_{\text{critical}}$ for an infinity observer only rely on the factor $\frac{\overline{g}_{\phi\phi}(r)}{g_{tt}(r)}$. In the Schwarzschild-like black hole of bumblebee gravity, the spacetime metric components $g_{tt}$ and $g_{\phi\phi}$ are the same with Schwarzschild black hole (the components $g_{rr}$ are different because of the presence of Lorentz-violating parameter $l$), so the unstable photon sphere $r=3M$ and black hole shadow radius $r_{sh}=b_{\text{critical}}=3\sqrt{3}M$ remain unchanged. 

\section{Equivalence between Our Geometric Approach and Conventional Approach \label{section7}}

In this section, we demonstrate that our extension of the geometric approach (for general spherically symmetric black holes) is completely equivalent to the conventional approach based on effective potentials of test particles moving in the gravitational field. In the conventional approach, photon spheres / circular photon orbits are obtained by analyzing the extreme points of effective potential $V_{\text{eff}}(r)$. Particularly, the unstable photon sphere $r=r_{\text{unstable}}$ corresponds to the local maximum of effective potential, and the stable photon sphere $r=r_{\text{stable}}$ is the local minimum of effective potential. 

The Lagrangian for test particles moving in a black hole spacetime can be defined through \cite{Carroll}
\begin{equation}
	\mathcal{L} = \frac{1}{2} \cdot T \cdot T
	            = \frac{1}{2} \cdot g_{\mu\nu} \frac{dx^{\mu}}{d\lambda} \frac{dx^{\mu}}{d\lambda} 
	            = \frac{\epsilon}{2} \label{Lagrangian}
\end{equation} 
where $T$ is the tangent vector of particle orbits, and $\lambda$ is an arbitrary affine parameter. Further, $\epsilon$ is a constant such that for massless particles $\epsilon=0$, and for massive particles $\epsilon=-1$. Hence, the variation of the action $S=\int_{A}^{B}\mathcal{L} \cdot d\lambda$ eventually gives the null (or timelike) geodesics in black hole spacetimes for massless particles (or massive particles). Here, we only analyze the photon motions and photon spheres, so we concentrate ourselves on the  massless particle case $\epsilon=0$.

Using the spherically symmetric black hole metric in equation (\ref{spacetime metric}) and the corresponding conserved quantities in equations (\ref{conserved energy}) and (\ref{conserved angular momnetum}), the following radial geodesic equation can be derived from the above Lagrangian (\ref{Lagrangian})
\begin{equation}
	-\frac{g_{tt}(r) \cdot g_{rr}(r)}{2} \cdot  \bigg(\frac{dr}{d\lambda}\bigg)^{2}+V_{\text{eff}}(r)
	= \frac{1}{2} \cdot E^{2}
\end{equation}
where $V_{\text{eff}}(r)$ is the effective potential of photon moving in black hole spacetime \cite{footnote4}
\begin{equation}
	V_{\text{eff}}(r) = \frac{L^{2}}{2} \cdot \bigg[ -\frac{g_{tt}(r)}{\overline{g}_{\phi\phi}(r)} \bigg]
\end{equation} 
Without loss of generality, we have used $\theta=\pi/2$ and constrained photon motions in the equatorial plane of black hole spacetimes. Furthermore, if one restricts the spacetime metric as $g_{tt}(r)=-f(r)$, $g_{rr}(r)=1/f(r)$, $g_{\theta\theta}(r)=r^{2}$ and $g_{\phi\phi}(r,\theta)=r^{2}\sin^{2}\theta$, then radial geodesic equation for photon reduces to
\begin{equation}
	\frac{1}{2}\cdot\bigg(\frac{dr}{d\lambda}\bigg)^{2} +\frac{f(r)}{2} \cdot \frac{L^{2}}{r^{2}}
	= \frac{1}{2} \cdot E^{2}
\end{equation}
which is exactly the radial equation (9) in reference \cite{Qiao}.

In the conventional approach, photon spheres / circular photon orbits $r=r_{ph}$ correspond to extreme points of effective potential \cite{Carroll,Hartle}
\begin{equation}
	\frac{dV_{\text{eff}}(r)}{dr} \bigg|_{r=r_{ph}} 
	= 
	\bigg[
	  \frac{L^{2}}{2} \cdot \frac{\partial}{\partial r} \bigg( -\frac{g_{tt}(r)}{\overline{g}_{\phi\phi}(r)} \bigg)
	\bigg]_{r=r_{ph}} 
	= 0  \label{extreme point equation0}
\end{equation}
and
\begin{eqnarray}
	&  & \frac{dr}{d\lambda} \bigg|_{r=r_{ph}} = 0 \nonumber
	\\
	& \Rightarrow & 
	V_{\text{eff}}(r=r_{ph}) = 
	\bigg[
	-\frac{L^{2}}{2} \cdot \frac{g_{tt}(r)}{\overline{g}_{\phi\phi}(r)} 
	\bigg]_{r=r_{ph}}
	= \frac{E^{2}}{2} 
	\label{photon sphere equation}
\end{eqnarray}
Note that the angular momentum $L$ in equation (\ref{extreme point equation0}) is a conserved quantity along photon orbits, which is independent of coordinate $r$. For photons moving in photon spheres / circular photon orbits, their conserved angular momentum $L = g_{\phi\phi}(r,\theta) \cdot \frac{d\phi}{d\lambda} = \overline{g}_{\phi\phi}(r) \cdot \frac{d\phi}{d\lambda}$ is non-vanishing. Therefore, we get the following relation from equation (\ref{extreme point equation0}) for photon spheres
\begin{eqnarray}
	\frac{\partial}{\partial r} \bigg( \frac{g_{tt}(r)}{\overline{g}_{\phi\phi}(r)} \bigg) \bigg|_{r=r_{ph}} = 0 \nonumber
	\\ 
	\frac{\partial}{\partial r} \bigg( \frac{\overline{g}_{\phi\phi}(r)}{g_{tt}(r)} \bigg) \bigg|_{r=r_{ph}} = 0  
	\label{extreme point equation00}
\end{eqnarray}
Recall the geodesic curvature condition of photon spheres in equation (\ref{geodesic cuurvature general}), it is clearly demonstrated that photon spheres obtained in our geometric approach and those in conventional approach are equivalent to each other. When $r=r_{ph}$ is the extreme point of the effective potential $V_{\text{eff}}(r)$, the geodesic curvature $\kappa_{g}$ along the sphere $r=r_{ph}$ vanishes automatically, which makes it to be a photon sphere / circular photon orbit for spherically symmetric black holes.

Next, we turn to the analysis of photon sphere stability. For the unstable photon sphere $r=r_{\text{unstable}}$, the effective potential of photon reaches its local maximum \cite{Hartle,Raffaelli2021}
\begin{widetext}
\begin{eqnarray} 
	\frac{d^{2}V_{\text{eff}}(r)}{dr^{2}} \bigg|_{r=r_{\text{unstable}}} 
	& = & 
	\bigg[
	  \frac{L^{2}}{2} \cdot \frac{\partial^{2}}{\partial r^{2}} \bigg( -\frac{g_{tt}(r)}{\overline{g}_{\phi\phi}(r)} \bigg)
	\bigg]_{r=r_{\text{unstable}}}  %\nonumber
	\\
	& = &
	\frac{L^{2}}{2} \cdot \frac{\partial}{\partial r} 
	\bigg[
	  \bigg( \frac{g_{tt}(r)}{\overline{g}_{\phi\phi}(r)} \bigg)^{2} \cdot 
	  \frac{\partial}{\partial r} \bigg( \frac{\overline{g}_{\phi\phi}(r)}{g_{tt}(r)} \bigg) 
	\bigg] 
	\bigg|_{r=r_{\text{unstable}}}  \nonumber
	\\
	& = & 
	\bigg\{
	  \frac{L^{2} \cdot g_{tt}(r)}{\overline{g}_{\phi\phi}(r)} 
	  \cdot \frac{\partial}{\partial r} \bigg( \frac{g_{tt}(r)}{\overline{g}_{\phi\phi}(r)} \bigg)
	  \cdot \frac{\partial}{\partial r} \bigg( \frac{\overline{g}_{\phi\phi}(r)}{g_{tt}(r)} \bigg)
	  +\frac{L^{2}}{2} \cdot 
	   \bigg( \frac{g_{tt}(r)}{\overline{g}_{\phi\phi}(r)} \bigg)^{2} 
	   \cdot \frac{\partial^{2}}{\partial r^{2}} 
	   \bigg( \frac{\overline{g}_{\phi\phi}(r)}{g_{tt}(r)} \bigg)
	\bigg\}_{r=r_{\text{unstable}}} \nonumber
	\\
	& = &
	\bigg\{ 
	  \frac{L^{2}}{2} \cdot \bigg( \frac{g_{tt}(r)}{\overline{g}_{\phi\phi}(r)} \bigg)^{2} \cdot 
	  \frac{\partial^{2}}{\partial r^{2}} \bigg( \frac{\overline{g}_{\phi\phi}(r)}{g_{tt}(r)} \bigg)
	\bigg\}_{r=r_{\text{unstable}}} < 0 
	\label{maximum point equation0}
\end{eqnarray}
In the derivation of last line, we have used the simplified extremum condition (\ref{extreme point equation00}) for unstable photon spheres $r=r_{\text{unstable}}=r_{ph}$. We also note that the factor $\frac{L^{2}}{2} \cdot \big( \frac{g_{tt}(r)}{\overline{g}_{\phi\phi}(r)} \big)^{2}$ is always positive for photon orbits, then the local maximum of photon effective potential in black hole spacetimes requires
\begin{equation}
	\frac{d^{2}V_{\text{eff}}(r)}{dr^{2}} \bigg|_{r=r_{\text{unstable}}} < 0
	\ \Rightarrow \ 
	\frac{\partial^{2}}{\partial r^{2}} \bigg( \frac{\overline{g}_{\phi\phi}(r)}{g_{tt}(r)} \bigg) 
	\bigg|_{r=r_{\text{unstable}}} < 0
\end{equation}

On the other hand, in our geometric approach to more general spherically symmetric black holes, unstable photon spheres indicate that the Gaussian curvature in two-dimensional optical geometry must be negative. Therefore, for the unstable photon sphere, we have the following inequality
%\begin{widetext}
\begin{eqnarray}
	\mathcal{K}(r=r_{\text{unstable}}) 
	& = & 
	\bigg\{
	  \frac{g_{tt}(r)}{\sqrt{g_{rr}(r)\cdot \overline{g}_{\phi\phi}(r)}}
      \cdot \frac{\partial}{\partial r}
      \bigg[
        \frac{g_{tt}(r)}{2\sqrt{g_{rr}(r)\cdot \overline{g}_{\phi\phi}(r)}}
        \cdot \frac{\partial}{\partial r} \bigg( \frac{\overline{g}_{\phi\phi}(r)}{g_{tt}(r)} \bigg)
      \bigg]
    \bigg\}_{r=r_{\text{unstable}}}  \nonumber
    \\
    & = & 
    \bigg\{
      \frac{g_{tt}(r)}{\sqrt{g_{rr}(r)\cdot \overline{g}_{\phi\phi}(r)}} 
      \cdot \frac{\partial}{\partial r}
      \bigg(
        \frac{g_{tt}(r)}{2\sqrt{g_{rr}(r)\cdot \overline{g}_{\phi\phi}(r)}}
      \bigg)
      \cdot \frac{\partial}{\partial r} \bigg( \frac{\overline{g}_{\phi\phi}(r)}{g_{tt}(r)} \bigg) 
    %\bigg\}_{r=r_{\text{unstable}}} \nonumber
    + %\bigg\{
        \frac{g_{tt}(r) \cdot g_{tt}(r)}{2g_{rr}(r)\cdot \overline{g}_{\phi\phi}(r)} 
        \cdot \frac{\partial^{2}}{\partial r^{2}} \bigg( \frac{\overline{g}_{\phi\phi}(r)}{g_{tt}(r)} \bigg) 
      \bigg\}_{r=r_{\text{unstable}}}  \nonumber
    \\
    & = & \frac{g_{tt}(r)\cdot g_{tt}(r)}{2g_{rr}(r)\cdot \overline{g}_{\phi\phi}(r)} 
          \cdot \frac{\partial^{2}}{\partial r^{2}} \bigg( \frac{\overline{g}_{\phi\phi}(r)}{g_{tt}(r)} \bigg) 
          \bigg|_{r=r_{\text{unstable}}} < 0
          \label{Gauss cuurvature general}
\end{eqnarray}
In this derivation, the simplified extremum condition (\ref{extreme point equation00}) is used in the last line. For a spherically symmetric black hole spacetime, $g_{rr}(r)>0$, $g_{\theta\theta}(r)>0$, $\overline{g}_{\phi\phi}(r)=g_{\phi\phi}(r,\theta=\pi/2)>0$ and $g_{tt}(r)<0$ hold outside the black hole horizons (and the optical geometry is usually defined outside the horizon), the Gaussian curvature condition for unstable photon spheres leads to
\begin{equation}
	\mathcal{K}(r=r_{\text{unstable}}) < 0
	\ \ \Leftrightarrow \ \ 
	\frac{\partial^{2}}{\partial r^{2}} \bigg( \frac{\overline{g}_{\phi\phi}(r)}{g_{tt}(r)} \bigg) 
	\bigg|_{r=r_{\text{unstable}}} < 0
\end{equation}
which is completely equivalent to the condition obtained from the conventional approach using effective potential of test particles. 

The similar conclusion hold for stable photon spheres. From the conventional approach, the stable photon spheres would correspond to the local minimum of effective potential. From a similar derivation, we can obtain the following relation for stable photon spheres
\begin{equation}
	\frac{d^{2}V_{\text{eff}}(r)}{dr^{2}} \bigg|_{r=r_{\text{stable}}} > 0
	\ \ \Leftrightarrow \ \ 
	\frac{\partial^{2}}{\partial r^{2}} \bigg( \frac{\overline{g}_{\phi\phi}(r)}{g_{tt}(r)} \bigg) 
	\bigg|_{r=r_{\text{stable}}} > 0
	\ \ \Leftrightarrow \ \ 
	\mathcal{K}(r=r_{\text{stable}}) > 0
\end{equation} 
which also shows the equivalence between the geometric approach and the conventional approach. %(based on effective potential of test particles).

\begin{table*}
	\caption{The distinguishing features and equivalence between the geometric approach and the conventional approach to photon spheres for general spherically symmetric black holes.}
	\label{table2}
	\vspace{2mm}
	\begin{ruledtabular}
		\begin{tabular}{lcccc}
			& Approach               & Geometric Approach & Conventional Approach &
			\\
			\hline
			& Geometry               & Optical Geometry of Black Hole Spacetime  & Spacetime Geometry &
			\\
			&                        & (Riemannian / Randers-Finsler Geometry) \footnote{This table summarizes the results on spherically symmetric black holes, in which the optical geometry is a Riemannian manifold. The rotational symmetric black holes, whose optical geometry is Randers-Finsler manifold, is left to on-going work.} & (Lorentz Geometry) &
			\\
			\hline
			& Key Quantities         & Gauss Curvature $\mathcal{K}(r)$ & Effective Potential $V_{\text{eff}}(r)$ &
			\\
			&                        & Geodesic Curvature $\kappa_{g}(r)$ &       
			\\
			\hline
			& Photon Sphere          & Conditions  & Conditions &
			\\
			\hline
			& photon sphere / circular photon orbit  & zero geodesic curvature & extreme point of effective potential &
			\\
			&                                        & $\kappa_{g}(r)=0$ & $\frac{dV_{\text{eff}}(r)}{dr}=0$ &
			\\
			& unstable photon sphere  & zero geodesic curvature and negative Gauss curvature \ \    & local maximum of effective potential &
			\\
			&                         & $\kappa_{g}(r)=0$ and $\mathcal{K}(r)<0$ & $\frac{dV_{\text{eff}}(r)}{dr}=0$ and $\frac{d^{2}V_{\text{eff}}(r)}{dr^{2}}<0$ &
			\\
			& stable photon sphere    & zero geodesic curvature and positive Gauss curvature \ \    & local minimum of effective potential &
			\\
			&                         & $\kappa_{g}(r)=0$ and $\mathcal{K}(r)>0$ & $\frac{dV_{\text{eff}}(r)}{dr}=0$ and $\frac{d^{2}V_{\text{eff}}(r)}{dr^{2}}>0$ &
			\\
		\end{tabular}
	\end{ruledtabular}
\end{table*}

Combining the calculations and discussions above, we can draw the conclusion for general spherically symmetric black holes. When $r=r_{\text{unstable}}$ is the unstable photon sphere near black holes, the effective potential of photon reaches its local maximum, meanwhile the Gaussian curvature must be negative. Conversely, when $r=r_{\text{stable}}$ is the stable photon sphere for spherically symmetric black holes, the effective potential of photon reaches its local minimum, and the Gaussian curvature at this position must be positive.
\begin{subequations} 
\begin{eqnarray}
	\frac{d^{2}V_{\text{eff}}(r)}{dr^{2}} \bigg|_{r=r_{\text{unstable}}} < 0 
	\ \ & \Leftrightarrow & \ \ 
	\mathcal{K}(r=r_{\text{unstable}}) < 0 %\nonumber
	\\
	\frac{d^{2}V_{\text{eff}}(r)}{dr^{2}} \bigg|_{r=r_{\text{stable}}} > 0 
    \ \ & \Leftrightarrow & \ \ 
    \mathcal{K}(r=r_{\text{stable}}) > 0
\end{eqnarray}
\end{subequations}
In this way, the equivalence between the geometric approach (based on Gaussian curvature and geodesic curvature in optical geometry) and the conventional approach (using effective potentials of test particles) for general spherically symmetric black holes is demonstrated. The distinguishing features of these two approaches and their equivalence are summarized in table \ref{table2}.

\end{widetext}

\section{Summary and Prospects \label{section8}}

The photon sphere and black hole shadow radius are important quantities in the theoretical and observational studies in black hole physics. In this work, we extend the geometric approach, which was proposed in a recent work by C. K. Qiao and M. Li to calculate the photon sphere and black hole shadow radius, %(in which the calculations are carried out only for a subclass of static and spherically symmetric black holes with spacetime metric $g_{tt}(r) \times g_{rr}(r) = -1$ and $g_{\theta\theta}(r)=r^{2}$), 
to more general spherically symmetric black holes. The conclusions in the present work are exactly the same with those obtained by Qiao and Li in reference \cite{Qiao}. Our approach is quite general, regardless of the particular metric form $g_{tt}$, $g_{rr}$ and $g_{\theta\theta}$. Our results indicate that the Gaussian curvature and geodesic curvature in optical geometry of black hole spacetime provide us new techniques to calculate the photon spheres and black hole shadows. Furthermore, we also demonstrate that the geometric approach in the present work is completely equivalent to the conventional approach based on the effective potential of test particles. This geometric approach may have a promising future in the investigations of photon spheres and black hole shadows, and the optical geometry of black holes could provide us with profound insights on gravity theory and black hole properties. 

It is also interesting that our geometric approach may provide additional evidence on one of the most famous issues in gravity theory: the nonlocal gravitational energy is tightly connected with the curvatures in Riemannian or Lorentz geometry. From this work, it is clearly indicated that the effective potential energy $V_{\text{eff}}(r)$ in spherically symmetric black hole spacetime is closely related to the Gaussian curvature in optical geometry. Historically, the connections between curvatures and gravitational energy were revealed by investigations of ADM mass and positive mass theorem \cite{Yau1979,Yau1981,Witten1981}, in which the scalar curvature and Gauss-Codazzi equation in differential geometry play important roles in elaborating these relations \cite{Lee2019}. 

In the near future, we will try to give a generalization of this geometric approach to stationary rotational black holes, whose optical geometry is a Randers-Finsler manifold. 

\begin{acknowledgments}
	The author acknowledge helpful discussions with Ming Li and Peng Wang during writing this manuscript. This work is supported by the Scientific and Technological Research Program of Chongqing Municipal Education Commission (Grant No. KJQN202201126), the Scientific Research Program of Chongqing Science and Technology Commission (the Chongqing “zhitongche” program for doctors, Grant No. CSTB2022BSXM-JCX0100), the Natural Science Foundation of Chongqing (Grant No. CSTB2022NSCQ-MSX0932), and the Scientific Research Foundation of Chongqing University of Technology (Grant No. 2020ZDZ027). The author thanks the anonymous referees of article \href{https://doi.org/10.1103/PhysRevD.106.L021501}{PRD {\bf 106}, L021501 (2022)}, whose valuable comments partially inspired us to generate this geometric approach for more general black holes. The author also thanks Xue-Rong Lan for kindly accompany and encouragements.
\end{acknowledgments}

\ \ \ 

\ \ \ 

{\bf Note:} During the uploading of this manuscript in the \href{https://arxiv.org/}{arXiv.org}, we find that some of the calculations and conclusions in the present work overlaps with a very recent work \href{https://doi.org/10.48550/arXiv.2207.14506}{arXiv:2207.14506[gr-qc]} (which is the reference \cite{Cunha2022}) submitted 3 days ago (on 29th July, 2022). In \href{https://doi.org/10.48550/arXiv.2207.14506}{arXiv:2207.14506[gr-qc]}, the authors discussed both timelike and null circular geodesics for general spherically symmetric black holes, but they did not discuss black hole shadows.

%\clearpage

\appendix
\section{Gaussian Curvature, Hadamard Theorem and the Stability of Photon Sphere \label{appendix1}}

\begin{figure}
	\includegraphics[width=0.525\textwidth]{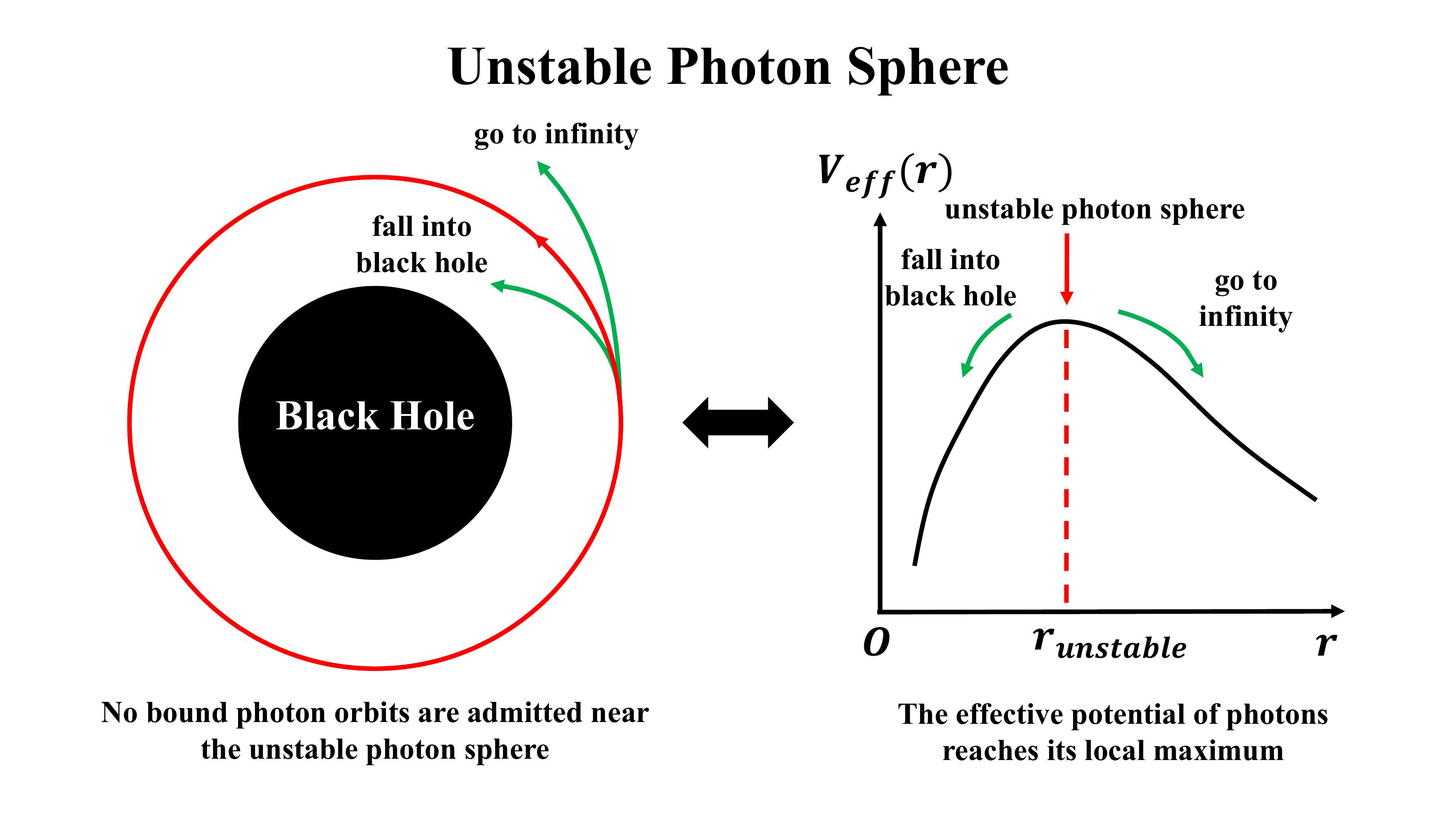}
	\includegraphics[width=0.525\textwidth]{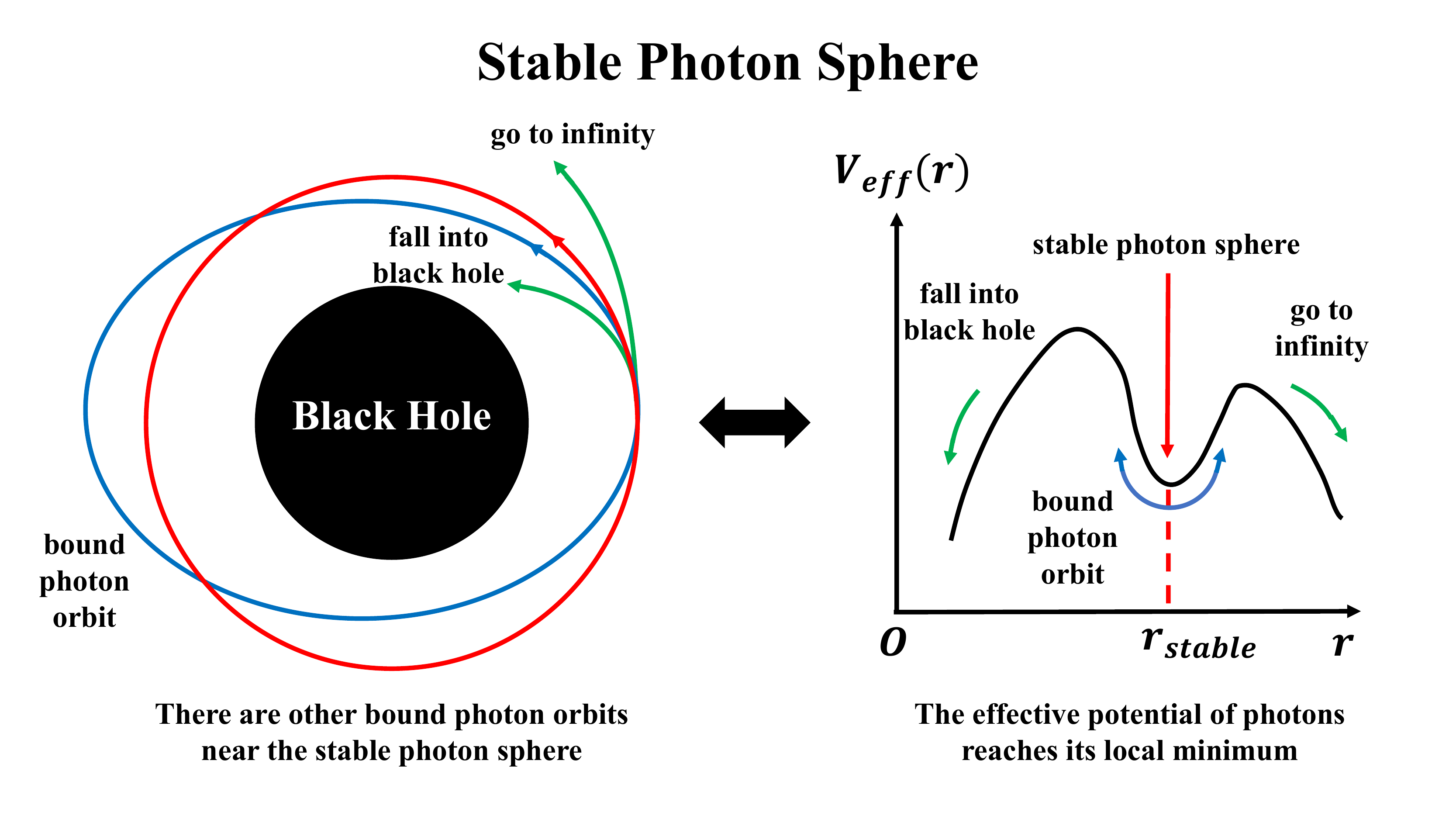}
	\caption{The stable and unstable photon spheres near black holes. The left part of this figure shows the stable and unstable photon spheres and other bound photon orbits nearby. The regions covered by black denote the interior of spherically symmetric black holes. The right part of this figure shows the behavior of photon effective potential $V_{\text{eff}}(r)$ near photon spheres.}
	\label{figure1}
\end{figure} 

\begin{figure}
	\includegraphics[width=0.525\textwidth]{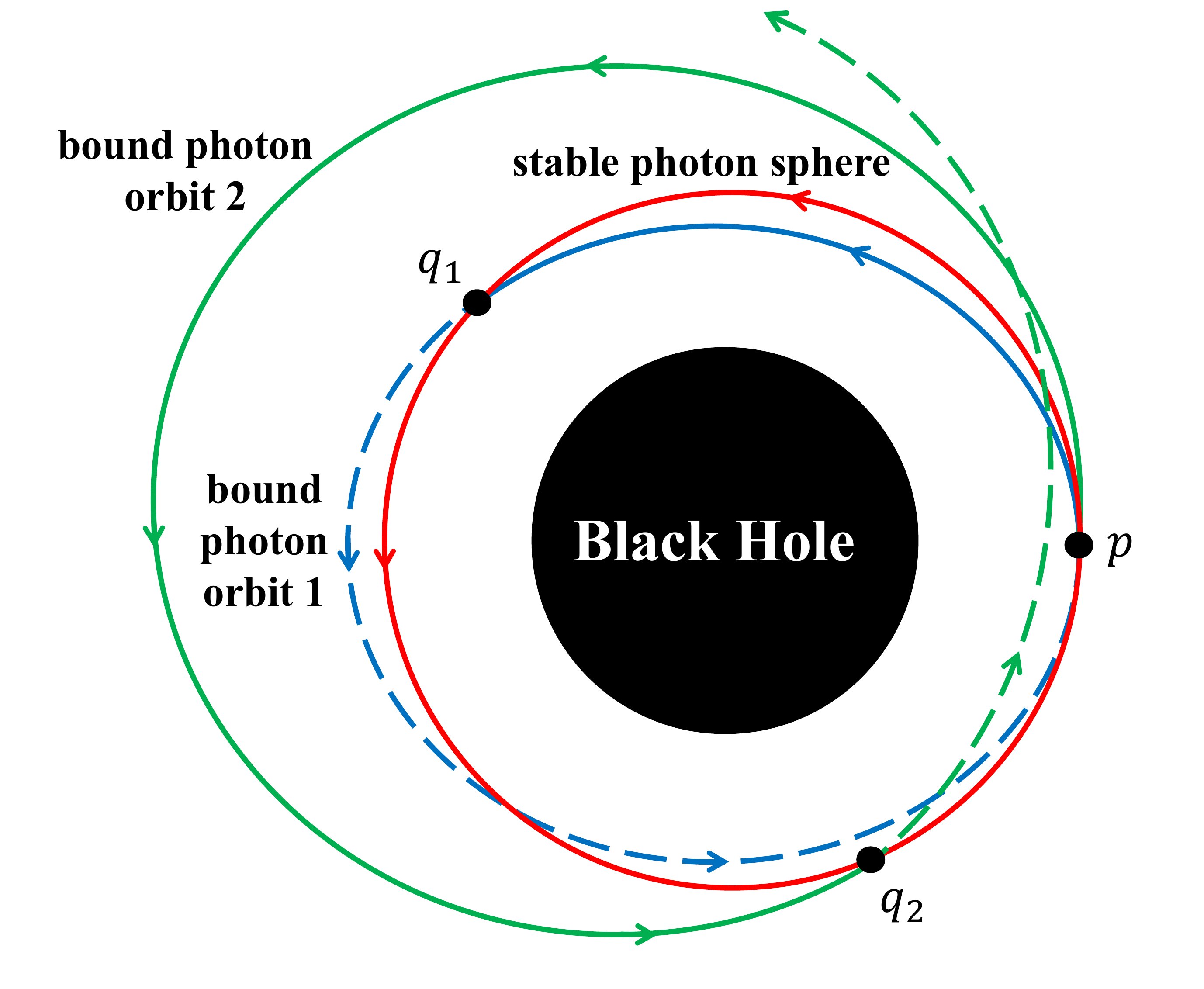}
	\caption{This figure shows several possible bound photon orbits near the stable photon sphere. These bound photon orbits could be obtained by a perturbation of the photon sphere at a particular point $p$, which may have different shapes. The bound photon orbit $1$ labeled in blue is a closed orbit, and the bound photon orbit $2$ labeled in green is an unclosed orbit (just like the Mercury procession around the Sun). In the equatorial plane of optical geometry, we can find another point $q$ (we denote $q_{i}$ for bound orbits $i=1,2$). When we consider photon orbits starting from $p$ ending with $q$, there are at least two different photon orbits (one is the stable photon sphere, and the other is a bound photon orbit) that can be continuously deformed to each other. Therefore, in the equatorial plane of optical geometry, there are two different geodesic curves from $p$ to $q$ belonging to the same homotopy class.}
	\label{figure2}
\end{figure}

In this appendix, we give a brief introduction of the relation between Gaussian curvature and the stability of photon spheres near black holes. The descriptions presented here are completely follow the arguments in reference \cite{Qiao}. According to the geometric approach developed by C. K. Qiao and M. Li, the stability of photon spheres are determined by the Gaussian curvature in the equatorial plane of optical geometry. The negative Gaussian curvature indicates the corresponding photon spheres are unstable, while the positive Gaussian curvature implies the corresponding photon spheres must be stable. The derivation of these arguments are based on the Hadamard theorem in surface theory, differential geometry and topology.

The photon spheres / circular photon orbits are important quantities of black holes. They are closely connected with particle motions, gravitational lensing and black hole shadows. Usually, the photon spheres near black holes can be classified into two categories: stable photon spheres and unstable photon spheres. They may exhibit significantly different features, especially for the geodesics nearby. For unstable photon spheres, when photon beams have a departure from the photon sphere, they would either fall into the black hole or go to infinity. In other words, no bound photon orbits are admitted near the unstable photon sphere. Conversely, there are many bound photon orbits near the stable photon sphere. These bound photon orbits may have rather different shapes (sometimes they probably be unclosed orbits, just like the Mercury procession around the Sun). 
The stable and unstable photon spheres are illustrated schematically in figure \ref{figure1}. 

An interesting question in black hole physics is how to distinguish these stable photon spheres from unstable photon spheres. The following Hadamard theorem in differential geometry would offer an answer to this question appropriately.
\begin{quote}
	\textbf{Hadamard Theorem:}
	For a two dimensional complete Riemannian manifold with  nonpositive Gaussian curvature, there is only one geodesic curve from $p$ to $q$ belong to the same homotopy class, and this geodesic curve minimizes the length in this homotopy class \cite{Berger2,Jost2011,footnote2}.
\end{quote} 

For stable photon spheres, there are bound photon orbits nearby, which may have different shapes. In the equatorial plane of optical geometry, we can always find two points $p$ and $q$ such that there are at least two geodesic curves (one is the stable photon sphere, the other is a bound photon orbit) belonging to the same homotopy class. Figure \ref{figure2} illustrates two possible bound photon orbits and the corresponding choice of points $p$ and $q$. In such cases, the Gaussian curvature in equatorial plane of optical geometry should be positive, otherwise it would violate the Hadamard theorem. On the contrary, for the unstable photon sphere, no bound photon orbits homotopic to this photon sphere exist. Then the unstable photon sphere itself forms the whole homotopy class, which corresponds to the negative Gaussian curvature in the Hadamard theorem (we assume the Gaussian curvature in the equatorial plane of optical geometry is nonzero, otherwise the black hole spacetime would be flat). According to properties of photon spheres and the Hadamard theorem, the following criterion using Gaussian curvature to distinguish the stable and unstable photon spheres is proposed \cite{Qiao}
\begin{eqnarray}
	\mathcal{K} < 0 & \Rightarrow & \text{The photon sphere $r=r_{ph}$ is unstable}  \nonumber \\
	\mathcal{K} > 0 & \Rightarrow & \text{The photon sphere $r=r_{ph}$ is stable} \nonumber
\end{eqnarray}

%\begin{acknowledgments}
%\end{acknowledgments}


%\bibliography{reference.bib}

\begin{thebibliography}{99}
\bibitem{footnote4}
	In some references, the radial geodesic equation for photons in spherically symmetric black holes are often written in a slightly different from
	\begin{equation}
		\frac{g_{rr}(r)}{2} \cdot \bigg( \frac{dr}{d\lambda} \bigg)^{2} + \widetilde{V}_{\text{eff}}(r) = 0
	\end{equation} 
    where the effective potential of photons are defined as
    \begin{equation}
    	\widetilde{V}_{\text{eff}}(r) = \frac{L^{2}}{2\overline{g}_{\phi\phi}(r)} + \frac{E^{2}}{2g_{tt}(r)} 
    \end{equation}
    where $\theta=\pi/2$ is used for equatorial plane, and $\overline{g}_{\phi\phi}(r)=g_{\phi\phi}(r,\theta=\pi/2)$ is the metric component restricted in this equatorial plane. From this radial geodesic equation, the photon spheres of black hole satisfies
    \begin{eqnarray}
    	&  & \frac{dr}{d\lambda} \bigg|_{r=r_{ph}} = 0 \nonumber
    	\\
    	& \Rightarrow & 
    	\widetilde{V}_{\text{eff}}(r=r_{ph}) = 
    	\bigg[
    	  \frac{L^{2}}{2\overline{g}_{\phi\phi}(r)} + \frac{E^{2}}{2g_{tt}(r)} 
    	\bigg]_{r=r_{ph}} = 0
    \end{eqnarray}
    and    
    \begin{equation}
    	\frac{d\widetilde{V}_{\text{eff}}(r)}{dr} \bigg|_{r=r_{ph}} 
    	= \frac{\partial}{\partial r}
    	\bigg[
    	  \frac{L^{2}}{2\overline{g}_{\phi\phi}(r)} + \frac{E^{2}}{2g_{tt}(r)} 
    	\bigg]_{r=r_{ph}} 
    	= 0  %\label{extreme point equation0}
    \end{equation}
    From a similar calculation, one can find that these conditions for photon sphere are consistent with our equations  (\ref{extreme point equation0}) and (\ref{photon sphere equation}) in the main-text.
\bibitem{footnote2}
    Although a black hole spacetime has singularities, its optical geometry is usually defied outside the horizons. Therefore, the optical geometry can be generally viewed as a geodesic complete manifold, which realizes the precondition in Hadamard theorem. Specifically, for static or stationary spacetimes, the time coordinate $t$ in spacetime geometry becomes arc length parameter in optical geometry, the endless and infinitely extendibility of time coordinate $t$ in photon orbits outside horizons implies the geodesic completeness of optical geometry.
\end{thebibliography}

\end{document}